# SINA - Smart Interoperability Architecture

## An architecture fostering the interoperability between smart building technology from different manufacturers and smart grid infrastructure to enable new business models for energy services

**Andreas Rumsch[1*], Christoph Imboden[2], Alberto Calatroni[1,] Martin Camenzind[1], Edith Birrer[1], Andrew Paice[1]**

[1]iHomeLab, Institute of Electrical Engineering

[2]Competence Center Power Economy, Institute for Innovation and Technology Management

Both at Lucerne University of Applied Sciences and Arts, Engineering and Architecture, Horw, Switzerland

**\* Correspondence:** Andreas Rumsch, andreas.rumsch@hslu.ch

**Abstract**

More and more household appliances connect to the Internet and exchange data freely. This is the foundation for true smart buildings. However, there is still no uniform communication technology available, which can connect all the different appliances from different vendors. Standards and protocols differ from manufacturer to manufacturer, which makes interoperability difficult or even impossible. Manufacturers cannot rely on a reference for the implementation and real estate developers and operators are reluctant to commit to a system until it is clear which one will prevail, providing investment security. A similar situation is evident in the area of smart grids and applies equally to the energy supply industry. This fragmentation ultimately leads to missed opportunities in terms of business models which could connect customers with service providers.

To overcome these limitations, we present a first draft of an architecture, which we call SINA - Smart Interoperability Architecture. SINA is based on existing decentralized infrastructure, which avoids creating a dependency of the market participants on an overpowering service provider. The core element of the technical solution is an open-source module integrated in the private clouds of the manufacturers, energy suppliers and service providers. The architecture particularly solves problems of data ownership, privacy and data security and avoids central administrative structures. It manages data access and transfer in a decentralized and distributed system, without needing dedicated hardware. SINA uses a blockchain and smart contracts to make sure that the pieces of information about which data are accessed, by whom they are accessed, how they are processed, and which monetary transactions take place are immutably stored and made available in a decentralized fashion. This allows providers to offer services to users in a transparent and trustworthy manner. Finally, SINA includes a matchmaking block which should automatically help service providers find potential customers and vice versa. This set of features makes SINA unique.

With the SINA project we are still at the very beginning. Therefore, this article is a position paper and does not show results from a running system. It describes proposals to tackle the questions that arise in the field of interoperability. We encourage readers to send us their thoughts and feedback.

# 1    Introduction

Due to an increase in new renewable energy sources and associated challenges, energy suppliers and distribution network operators start to create intelligent networks and automate their processes. In addition, more and more consumer products can connect to the Internet and exchange data freely, many of them with remote control capabilities already enabled.

The growth of connected technologies with automatic and remote-control possibilities may be observed at the grid and building level, giving rise to two independent domains – smart grids and smart buildings. Exploiting these two application domains in combination provides a huge advantage to enable the intelligent shifting of residential loads to increase grid stability. However, realizing such intelligent control mechanisms requires communication across the boundaries between smart grid and smart buildings. Due to the lack of consistent standards, a large number of protocols and proprietary solutions have emerged which renders interoperability and the creation of services that cross this boundary difficult or even impossible. Manufacturers cannot rely on a reference for the implementation and real estate developers and operators will not commit to a system until it is clear which one will prevail. This issue also complicates the development of business models for service providers.

Ongoing initiatives rely on the introduction of decentralized infrastructure such as field gateways and data concentrators. Such equipment is rarely acceptable for home appliance manufacturers, since they operate in a price sensitive market. Manufacturers prohibit any additional decentral hardware cost if not deemed absolutely necessary. This also applies to any additional decentral installation and maintenance cost.

Thus, a reasonable system architecture which overcomes these issues must build on existing infrastructure such as smart meters and smart appliances, since manufacturers have them already integrated in their private networks, providing monitoring and remote-control services within their own domains. In order to develop this architecture, we held a series of discussion and workshops with stakeholders from the smart grid and smart building ecosystems, identifying pains, gains and needs for the intelligent integration of all relevant systems at the grid and building level. The result of these discussions is the Smart Interoperability Architecture (SINA). In the remainder of this document, we use SINA to denote both the architecture and the working group who develops it.

The architecture deals with the following important features:

- Solving the problems of data ownership and data security in general.
- Only the data owner manages data access.
- Avoiding technical central administrative structures.
- Managing data access and data transfer in a decentralized and distributed system.
- Automating the matching between potential service providers and potential customers.

In this paper, we present a first draft of the architecture addressing data ownership, data security and management of data access across the boundary of smart grid and smart buildings, based on a decentralized, distributed approach. In a working group, several industrial companies, utilities and associations together with academia have elaborated requirements to an interoperability architecture as well as use cases that have to be realizable with SINA. In section 2 we illustrate the state of the art and why SINA goes beyond it. Section 3 contains the results of the requirements engineering as identified by the working group. Section 4 then describes the proposed architecture, comprising workflows and building blocks. Section 5 shows how existing solutions integrate into SINA. In section 6, we give an outlook on how we plan to proceed with the concrete implementation of SINA, including test and verification procedures.

# 2    State of the art

From a technical point of view, SINA fosters new business models by connecting devices from the domains smart building and smart grid. We therefore regard SINA as an architecture for the Internet of Things (IoT). In the following, we refer to several established IoT architectures and one smart grid architecture. We illustrate in which respect SINA differs or goes beyond the existing state of the art.

## 2.1    IoT architectures

The European Telecommunication Standards Institute proposed the oneM2M architecture for heterogeneous IoT systems [1], [2], which describes how things are connected via a network layer to the application in the application layer. The oneM2M architecture seeks to connect devices from different manufacturers to the cloud to make them available for applications. For example, a BACnet-automated air conditioning system should connect to temperature sensors that communicate with LoRa. The BACnet system and LoRa run on completely different systems and there is no natural connection point. oneM2M architecture addresses this problem with its horizontal framework and RESTful APIs. This enables the LoRa system to interact with the building automation system via an IoT network. oneM2M solves technical problems regarding communications but does not provide functionalities such as matchmaking.

In 2014, the IoTWF Architectural Committee published a reference model for an architecture with seven layers [3]. This reference model provides a clear and simplified perspective on IoT and includes edge computing, data storage and access. Each of the seven layers contains specific functions. Security is of great relevance throughout the model. This reference model tackles the issue of interoperability by a layered approach, where the communication takes place at the data abstraction layer. SINA tackles the interoperability at different levels, depending whether we need to achieve communication on the semantic level (through an ontology) or on a lower level, where we employ adapters.

A scalable and self-configuring peer-to-peer-based architecture for a large-scale IoT network has been proposed in [4]. The architecture aims to provide automated service and resource discovery mechanisms without human intervention. The architecture is based on an IoT Gateway which takes care of connecting heterogeneous items. The members of the SINA working group expressed their reluctance toward such a solution and SINA will rather use adapters to connect the manufacturers' clouds directly.

Z. Qin et al. [5] designed a software-defined network (SDN)-based architecture for the IoT with the objective of providing high-level quality of service (QoS) to the different IoT tasks in heterogeneous wireless network environments like electric vehicles, electric charging sites and smart grid infrastructure. The SDN-based architecture implements semantics (ontologies and rules) to match abstract functions with lower-level data communication protocols, offering a first step towards matchmaking. However, SINA aims to push the matchmaking up to the business level, where customers, service providers and manufacturers are automatically proposed scenarios which could lead to a contract.

In [6], the authors proposed a new IoT architecture called 3G-PLC where they combined power line communication (PLC) and third generation (3G) network to increase the scalability. This approach is quite specifically focusing on the communication layer and requires extra hardware, which is not the direction which we need for SINA.

In [7], the authors studied mobile phones as spontaneous gateways for wireless sensor networks in IoT systems. By using name-based future internet architecture (FIA) called MobilityFirst, they showed that many challenges associated with mobile phone gateways can be addressed. While this is an interesting approach, using phones as spontaneous gateways is not always possible or desirable, and does not cover all the use cases that we envision with SINA.

J. Zhou et al. [8] presented an IoT-enabled smart home scenario to analyse the IoT application requirements. The authors proposed the CloudThings architecture which accommodates platform as a service (PaaS), software as a service (SaaS) and infrastructure as a service (IaaS) to accelerate IoT applications. The CloudThings architecture is focusing on the smart home domain and lacks automatic mechanisms to interact with the smart grid on the data, semantic and business level.

Open-source architectures like the Distributed Services Architecture (DSA) are also available. The DSA is an IoT platform that facilitates device inter-communication. This network topology consists of multiple Distributed Service Links running on edge devices connected to a tiered hierarchy of brokers. This allows the system as a whole to be scalable, resilient to failure and to take advantage of all computing resources available to it from the edge, the datacenter, the cloud and everything in between (http://iot-dsa.org/). The DSA is quite close in its spirit to what SINA wants to achieve, but is lacking the ontological reasoning and the smart contract execution which will enable matchmaking and provide the necessary level of trust.

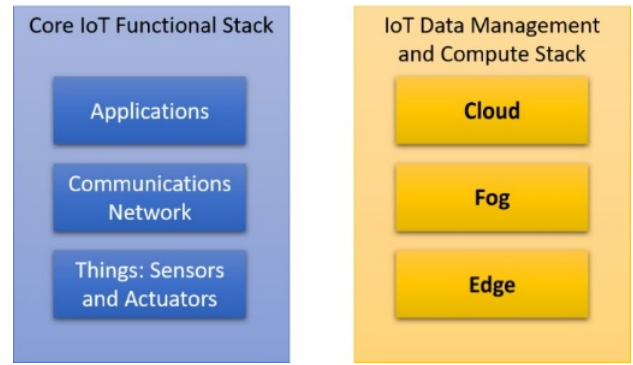

*Figure 1: Simplified IoT*

The authors in [9] describe a simplified architecture (Figure 1) representing the commons of most IoT architectures: they divide the IoT into layers, enabling the more or less independent development of technologies and standards on the individual layers. They also assume that the IoT endpoints are connected to a network that transports the data that is ultimately used by applications. In the Core IoT Functional Stack the resulting architecture describes how things are connected via the different layers known from other architectures to the applications. An additional point of view, the data processing, shows the IoT Data Management Stack. Here, the layers cloud, fog and edge are differentiated. Also in this architecture, matchmaking and decentralised smart contracts are missing.

Fog computing is one good example of a distributed architecture where edge devices (like routers, switches, gateways, or other small computers) carry out a substantial amount of computation, storage and communication locally while routing the information over the internet backbone. Fog computing therefore bridges the gap between cloud and end devices (IoT-nodes). Fog nodes reside close to IoT nodes, which in turn reduces latency.

Yousefpour et al. [10] provides a classification of typical computing paradigm where fog computing is a subset of the traditional cloud computing. Other distributed computing paradigms are mobile computing, edge computing, mobile cloud computing, mobile ad hoc cloud computing and mist computing. SINA strives to seamlessly integrate all these important and complementary paradigms.

## 2.2    Smart grid architectures

Among the architectures developed in the context of smart grids, we discuss briefly the Smart Grid Architecture Model (SGAM) [11]. The SGAM is a comprehensive model covering interoperability and business cases over different layers, focusing on the electrical grid. The reference description states explicitly that home automation and buildings are not in the scope of SGAM. On the contrary, SINA aims to enable the interplay of smart grid services and the appliances present in each building, such as heating, ventilation and air conditioning, boilers, heat pumps and electric vehicles.

Another aspect where SINA goes beyond SGAM is the definition of special building blocks and operational procedures. The most prominent ones are the matchmaking algorithm / phase and the use of smart contracts to execute code and data sharing in a transparent and immutable way. It is also worth mentioning that SGAM is meant to guide standardization bodies and technical groups into modelling their processes along the lines of the architecture described in SGAM. SINA is rather an IoT architecture enhanced by procedures which can be directly implemented as a middleware in existing decentral hardware. A last difference between SGAM and SINA is in the data modelling approach. SGAM suggests a top-down approach and explicitly recommends to avoid data translators which increase the complexity, whereas SINA builds explicitly on top of existing protocols and systems in a bottom-up fashion. We deem this necessary to maximize the acceptance and reuse of existing building blocks and to reduce the extra modelling work to a minimum.

SGAM defines cross-cutting issues which affect all model layers. These issues are relevant also for SINA and similar considerations were made within the working group while designing the SINA concept.

## 2.3    SINA compared to existing architectures

While the specific differences between SINA and each of the architectures are outlined above, the most prominent distinction between SINA and the other IoT architectures is that SINA defines a matchmaking phase, executed by matchmaking algorithms, and a decentral infrastructure coupled with smart contracts. The former makes it easier for service providers and/or manufacturers to find potential customers, generating value and useful services; the latter makes sure that the contracts are immutable and executed on an infrastructure which is not dependent on a single vendor, avoiding the emergence of a dominant position.

The second aspect handled by SINA is the following: Established IoT architectures describe how devices connect to the cloud defining protocols and communication. The aim of the architectures is that devices from different manufacturers connect to the system built upon the respective architecture. Therefore, a manufacturer has to decide which architecture it wants to support. To connect devices from different manufacturers, the architectures imply that every device implements the same protocol and that they connect to the same cloud server. Often manufacturers do not want to share data with other clouds, particularly for security relevant devices such as locking systems or monitoring systems. SINA overcomes this situation by providing an architecture that enables interoperability across the boundaries of systems built on different IoT architectures. SINA allows to make existing building blocks interoperable instead of having to model the whole semantic, communication and data layers from ground up. Only where no appropriate approach is available, we will create it. An analysis of existing protocols, technologies, associations, and their relevance for SINA is available for the interested reader in the Appendix (chapter 10)

## 3    Vision, requirements and use cases

In focus group discussions with several energy supply and industrial companies, we identified pains and gains concerning the main challenge, interoperability. The technical and commercial potential of a smart decentralized infrastructure in the framework of the smart grid is promising and all the participants of the focus group see the advantages of an architecture that enables interoperability and of SINA in particular. In general, companies want to build new business models based on interoperability between smart buildings and the smart grid, but there are some issues hindering them to do so:

- It is difficult to connect systems such as the Smart Meter infrastructure with the building automation system because of the lack of well-defined interfaces.
- Even if protocols are in place, it is not always possible to use them because of incompatibilities between implementations.
- Data from important devices such as heat-pumps and boilers are not available and there is no standardized way of controlling them.

## 3.1    Vision and requirements

The vision of the focus group members is manifold. As a result of the discussions, we summarize the most important aspects of the vision as follows:

- SINA builds on the existing hardware infrastructure (manufacturer cloud) and does this in a decentralized way.
- A service provider can use the provided communication infrastructure, to enable new and innovative services.
- The system architecture should be simple and offer service providers the greatest possible flexibility in implementing the services.
- A unified system architecture, protocols and communication interfaces should allow joining arbitrary components together.
- Multiple services are able to access the same data, supporting the sharing of expensive communication infrastructure between services.
- Existing virtual private networks of appliance manufacturers should continue to function as of today, assuring the support of the manufacturers for the SINA open access solution.
- Negative impacts on the business of appliance manufacturers must be avoided as much as possible.
- The management of data access is easy even for untrained data owners, and responsibility for data access management is with the data owners.
- Additional overhead because of data security and privacy should be minimal.
- Sharing of data and functionality should provide benefits for all participants (customer, manufacturer and service provider).
- A dependence on a single market participant is avoided.

To be able to transform this vision into reality, we set out to find the important requirements, which we divide into the following four groups: Governance, Service, Architecture and Security

### Governance

SINA should develop a set of governance rules that guarantee maintenance, operation, and further development of the solution, while ensuring that sovereignty over the solution is exercised by an organization that acts for the benefit of the general public and not for profit.

### Service

Subscription to services, for both data provision as well as data request, shall consider quality requirements such as real-time capability, resolution, etc.

### Architecture

- SINA must provide reference implementations to make it easy to integrate the SINA interface in any existing infrastructure – be it a manufacturer's private network and own server or a field device offering direct SINA access.
- The SINA architecture shall be made available as open source, reducing the risk for the energy service providers and manufacturers of a hold-up or lock-in situation.
- The SINA architecture shall, as much as reasonably possible, utilize existing standards and solutions, in order to increase the acceptance of the solution among the different parties and, at the same time, reduce the implementation cost and risk.

### Data security

- Considering data security, ownership and privacy from the start up is one of the most important aspects of SINA.
- Only the data owner – which in many cases may be a non-technical person - can provide access to the data. This includes, among other things, that the data owner determines which service provider may evaluate and use his data or control devices and how.
- The SINA system must prevent data misuse by employing strong authentication, encryption and access control mechanisms.

## 3.2    Challenges

### Open standard

Open standards and protocols are the foundation of the SINA system and increase the acceptance among industry partners involved. The definition of new standards and protocols is difficult because it requires the agreement of all affected parties on a final set of features. To simplify this process, we seek to integrate existing standards as much as possible, which broadens the circle of interested parties and helps SINA in gaining support.

### Security

The data owner alone must be able to manage the access to data points, which complies with today's societal expectations on security. The system must therefore offer a way to manage access control in a way non-technical people can handle and understand. This also means that the manufacturer and technical provider of the data access can delegate responsibilities, trusting in the security of the SINA system.

**Privacy**

Data owners have the right to decide which service can access the data and for how long. Data owners also have the right to revoke all rights to process and analyze data. SINA must always provide ways to ensure the privacy of all users. This also includes ways to enforce the deletion of previously distributed data used to provide the respective services by any service provider.

**Computational overhead**

Because SINA is a decentralized, distributed framework and should be built upon existing cloud infrastructure of appliance manufacturers, we expect additional computing capacity needed to implement the required functionality in the SINA modules. This extra computing capacity can be explained by the increased computational complexity for security and adaption to existing standards and protocols. Additionally, service providers might increase the frequency and amount of data transferred. This increase in traffic must be carefully analyzed and balanced.

### 3.3 Use cases

Accessing data from the energy infrastructure can enable new and innovative services, for example in the field of elderly care. Likewise, energy management systems get access to additional data points from different sensors, such as presence control, access control, etc. Having access to additional data points is considerably improving the quality of state analysis and load forecasting. The full potential of advanced energy management algorithms is realizable.

SINA provides the possibility to create new services based on the interconnection of appliances and the integration of the smart grid. Some examples are:

- Smart building services
- Local monitoring & management
- Remote maintenance
- Facility management
- Security services
- Ambient assisted living
- Health care services
- Smart grid services
- E-contracting
- Alternative pricing models
- Peak shaving
- Balancing services

- Voltage regulation
- Self-balancing
- Energy portfolio optimization
- Grid planning
- Congestion management

### 3.4 Selected sample use case: "Ancillary Services"

In this section we illustrate in more depth a concrete use case which would be enabled by using SINA. The electrification of the transportation sector is progressing at an increasing pace, posing challenges to the electrical grid, but also offering important opportunities for several stakeholders. In fact, having a fleet of electric vehicles (EVs), which stay most of the time attached to the electric grid through charging points, offers the possibility of using the large batteries embedded in the EVs as a collective energy reserve for ancillary services. These services exploit so-called flexibilities, such as batteries providing stored energy which can be used as a backup to guarantee the stability of the electrical grid whenever there is excess in electricity demand compared to generation. To explain the use case and how SINA would support it, let us start by outlining the stakeholders:

- Distribution system operators (DSOs), or electrical network providers.
- Flexibility aggregators – companies providing backup ancillary services to DSOs.
- Charging point operators (CPOs).
- Car sharing companies.
- EV owners living in buildings equipped with a building management system (BMS).

The use case consists in having EV owners and car sharing companies using bidirectional chargers, therefore selling electricity back to the DSO whenever the grid operator needs urgent access to a primary electricity reserve, i.e., one which has to be made available quite quickly. At the same time, charging point aggregators, flexibility aggregators participate in the value chain by providing the control algorithms which can decide when to activate which flexibility. EV owners are remunerated with a lucrative price for each kWh of energy sold back. DSOs covering short-term electricity needs by flexibilities can avoid costly upgrades of the grid infrastructure. We will explain how SINA supports this use case in chapter 5, after having described in detail the SINA architectural blocks and operation.

## 4  The SINA architecture in detail

### 4.1  General description

As we described in chapter 2, most manufacturers have their own cloud solution, each of them representing an IoT system. Not all manufacturers use the same architecture or the same standards and protocols for the IoT system. Nor they do not want to change their existing systems to enable interoperability with a uniform IoT architecture. Indeed, this option may not be possible at all, because the requirements for the IoT infrastructure in the smart building sector are different from those in the smart grid sector, which is ultimately a critical infrastructure.

Our idea is to build an ecosystem, where manufacturers, service providers and customers – acting as data owners – are connected with each other using the SINA architecture (see Figure 2). A customer owns devices from various manufacturers. The devices typically connect to the manufacturer's cloud. The customer grants access to data and to remote control functions from those devices he wants to provide to a service provider. The service provider then in turn can offer his services based on the provided data and control options.

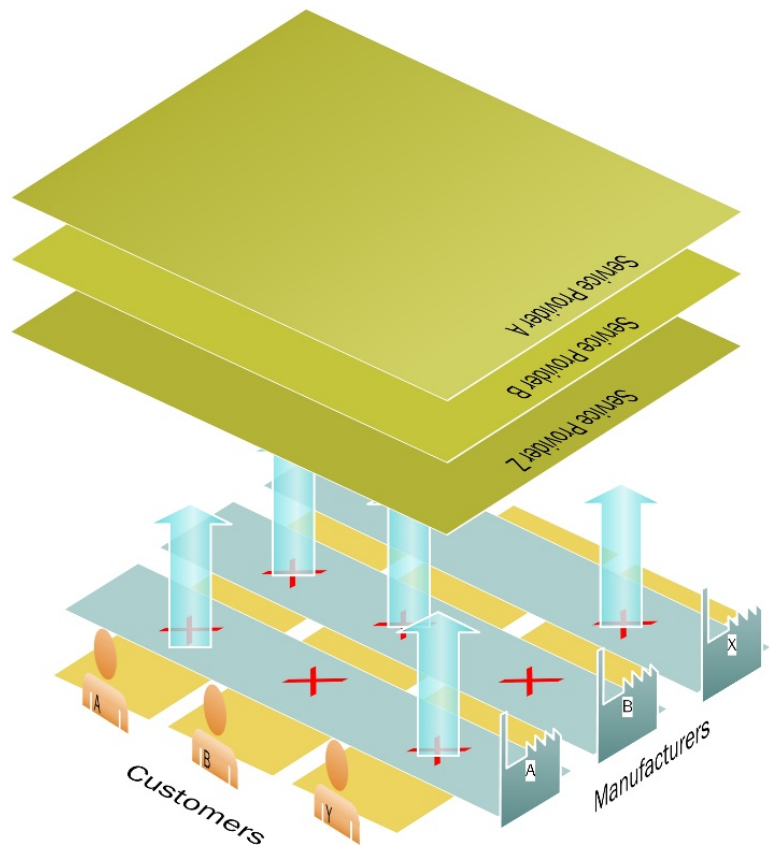

*Figure 2: SINA Ecosystem with customers, manufacturers and service providers participating in the SINA ecosystem.*

A more detailed view of the architectural design is shown in Figure 3. The customer connects his or her devices to the manufacturer's cloud via Internet using their own network infrastructure. Devices from different manufacturers are thus connected to different cloud solutions. In turn, the devices of all customers from one manufacturer are integrated into the manufacturer's cloud solutions. The manufacturer makes the data and control options that customers have released available to the service providers via a SINA adapter. The service providers in turn request the data required for the service from the SINA system via APIs. The customer agrees directly with the service provider on which data the service provider will receive and which control options it can use by means of electronic (smart) contracts. Further, data is not stored centrally but is exchanged directly between the manufacturer's cloud and a service provider (Figure 3). Additionally, there is no provision for centralised management of the system. SINA is a distributed, decentralized system.

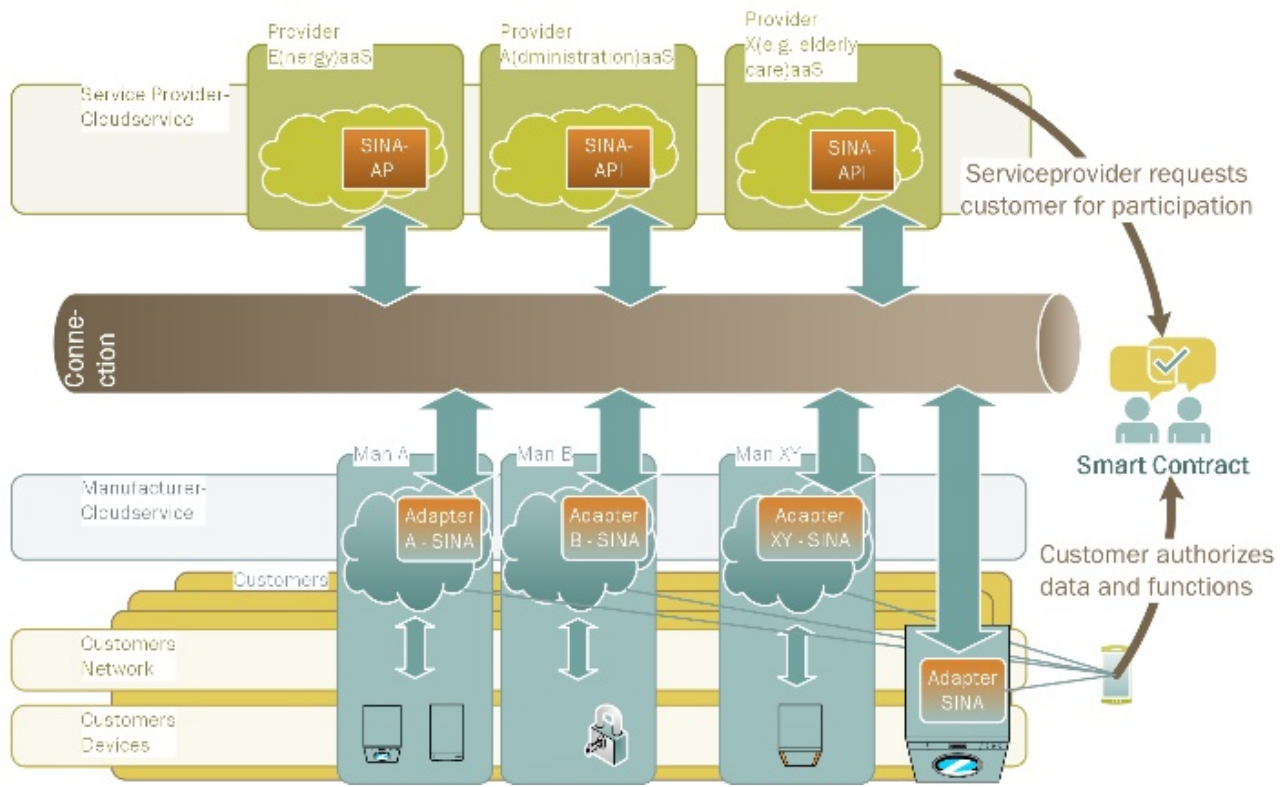

*Figure 3: Proposed architectural approach for SINA.*

The most important functions of the SINA system are ensuring data security, granting privacy, authentication, trust management and matchmaking (Figure 4). The functionalities are described in the following sections. We first describe the building blocks of SINA. We then explain how SINA works, distinguishing between the preparation phase, in which customers and service providers meet and agree on the service to be provided, and the operational phase, in which the service is provided.

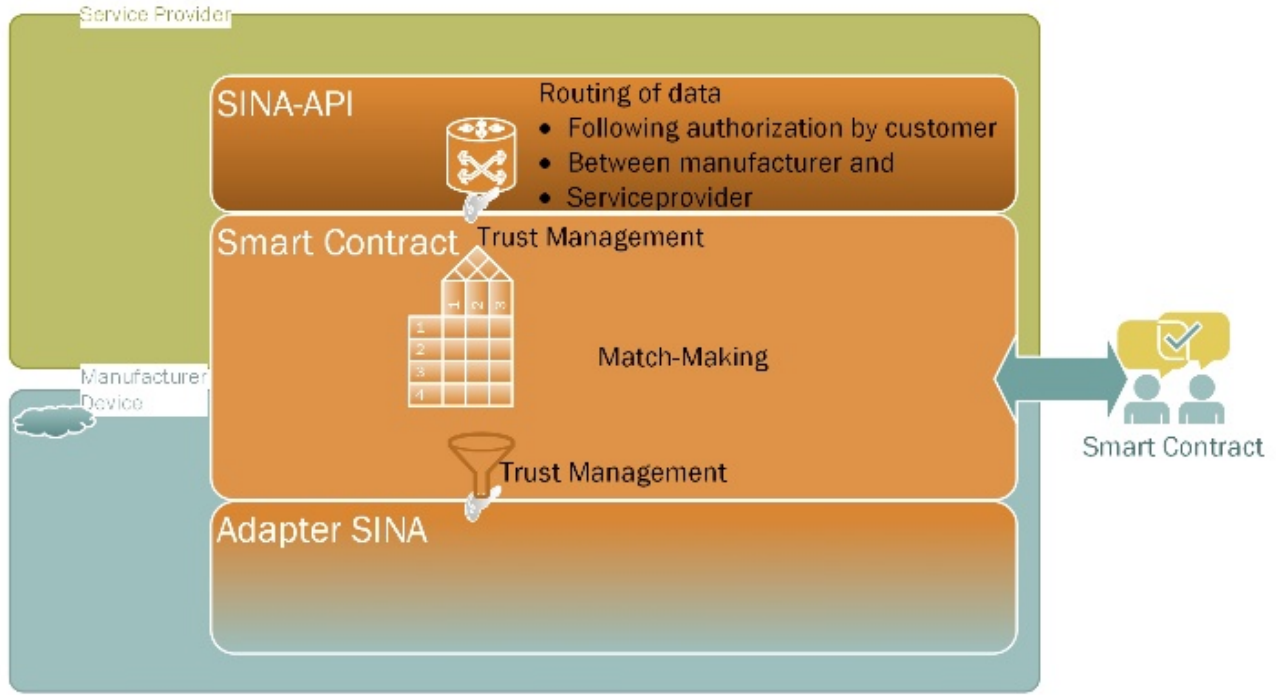

*Figure 4: Proposed approach for distributed functionality.*

## 4.2    Building blocks

Manufacturers and service providers must integrate the corresponding modules to participate in the SINA ecosystem. These modules consist of several building blocks. Figure 5 shows the building blocks for a manufacturer, Figure 6 those for a service provider. Descriptions of the building blocks follow in the next sections.

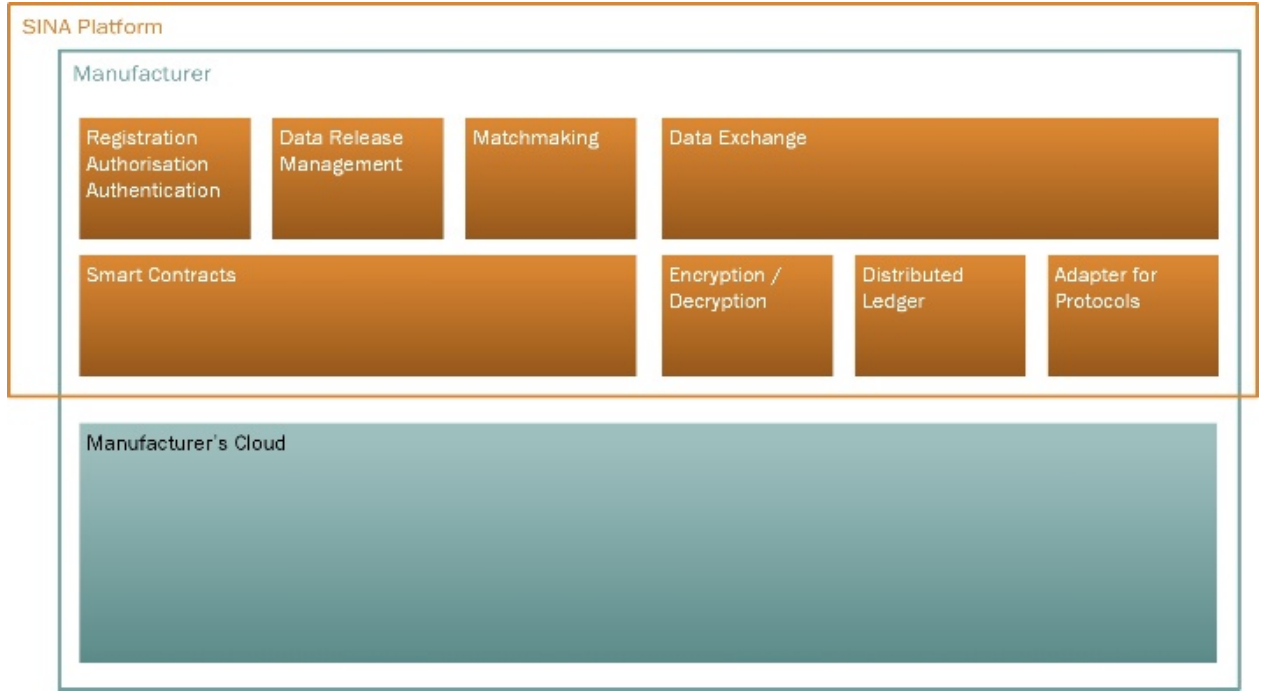

*Figure 5: Building blocks of a SINA integration for manufacturers.*

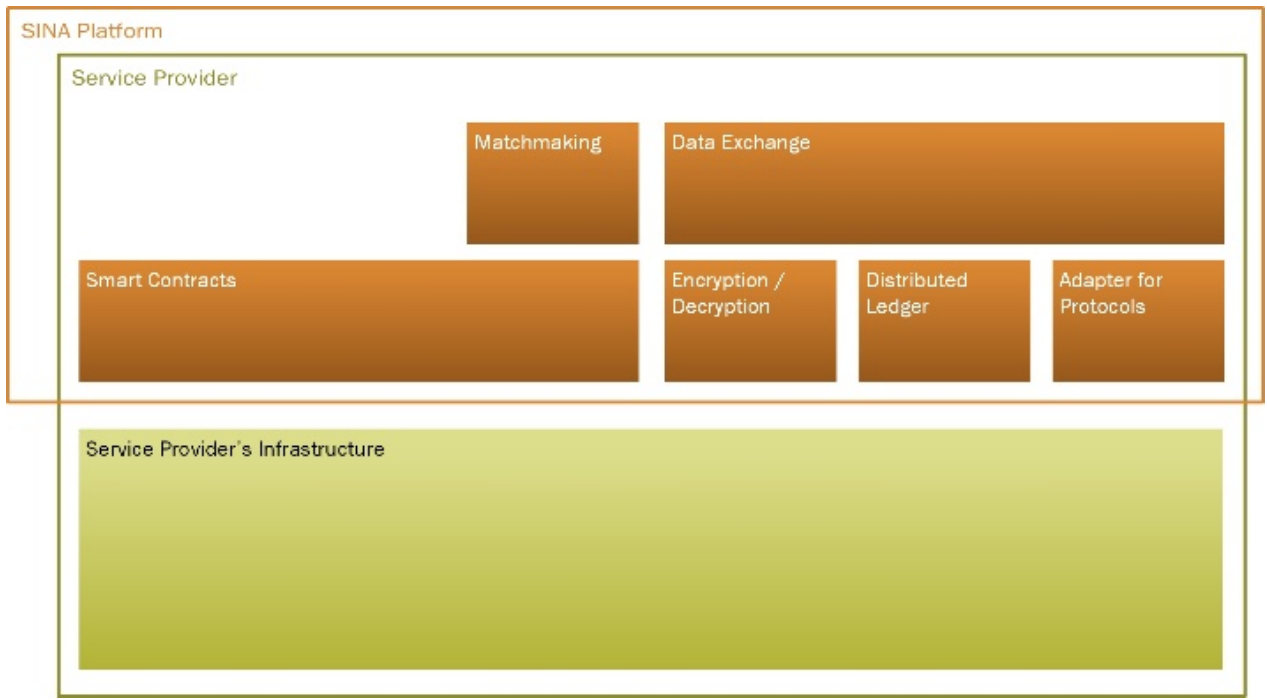

*Figure 6: Building blocks of a SINA integration for service providers.*

### 4.2.1 Registration, authorization, authentication

To participate in the SINA ecosystem, the identity of people and devices must be proven by a corresponding registration process. This process is handled by this building block. Due to the distributed nature of SINA, the registration of customers can be handled at every instance at any manufacturer. A customer that has registered with one manufacturer is identified for the whole SINA ecosystem. The customer is not obliged to register multiple times. This implies that SINA further needs a registration process for manufacturers.

Devices typically are identified by a serial number by the corresponding manufacturer. Through the registration process, the manufacturer guarantees for the identity of the device in the whole SINA ecosystem. A device has not to register at multiple manufacturers.

### 4.2.2 Data release management

Each manufacturer must manage the release of customer's data. The customer is responsible to define which data he wants to release. The release of data is controlled by an app where the customer can either select individual data or a profile provided by the manufacturer. Profiles contain a predefined set of data optimized for commonly used services.

The manufacturer's responsibility is to guarantee that only data and control options where the customer has granted access to is provided to selected service providers. This is one of the main functionalities of the SINA module.

### 4.2.3 Matchmaking

One of the challenges in SINA is to link service providers with suitable customers ready to share the required data. Where customers define which data and which control access they want to release, service providers provide a definition of data they need to perform the service. The definition comprises the type and quality of data. The type of data includes attributes such as temperature, energy consumption, state etc. of appliances, but also metadata such as information on the place of residence or eventually the person himself. The quality of the data defines accuracy, temporal resolution, real-time capability, etc.

The matchmaking unit searches for customers fulfilling a service provider's requirement or for service providers being able to work with the data a customer has granted access for. The matchmaking unit operates in the distributed system of SINA and is realized through ontological reasoning or search algorithms able to query distributed databases.

### 4.2.4 Smart contracts

When a customer and a service provider agree on a service, they conclude a smart contract. The service is then provided using the executable code in the smart contract. The smart contract needs access to external data, potentially from different manufacturers, which is provided through the SINA system. The use of smart contracts is an important prerequisite for the trust in the system, since every stakeholder knows that the kind of data, services and remuneration are bundled within an immutable smart contract. The immutability and traceability are guaranteed by using a blockchain / distributed ledger (see 4.2.7 below).

### 4.2.5 Data exchange

This building block implements all functionality to exchange data between the manufacturer's cloud and the service providers. Additionally, if data from different manufacturers are involved, this building block at service provider's side coordinates the exchange of data.

### 4.2.6 Encryption / decryption

The building block implements encryption and decryption of data using established encryption technologies.

### 4.2.7 Distributed ledger technologies / blockchain

This building block provides an essential functionality to make processes inalterable and traceable, thereby contributing to the trust management of the whole system. Mainly two use cases will rely on the functionality: The execution of smart contracts, and the exchange of certain data. Not every data transaction has to be inalterable and traceable. These properties apply only to use cases where valuables are exchanged with help of SINA. An example is energy trading where the traded energy represents a certain value. The exchange of temperature data, for example, does not fall into this category.

### 4.2.8 Adapter for protocols

To enable SINA to be compatible with existing protocols, this building block adapts the manufacturer's or service provider's cloud to the respective protocol. This means that protocols are translated from and to the protocols used by the manufacturer or service provider. The architecture will allow to implement several adapters for protocols in parallel. This makes SINA compatible with a variety of protocols.

### 4.3 Preparation phase

The deployment of SINA starts with a preparation phase. To ensure a high level of trustworthiness, SINA implements a trust management. One element of the trust management is to ensure the identity of participating customers, customer's devices, service providers and manufacturers. To gain access to a service in the SINA ecosystem, customers must first register themselves and their devices in the SINA system, Figure 7. Because of the decentralized nature of the solution, this can be done on any cloud system which implements the SINA module. Through the registration, the identity of people and devices is ensured. As far as possible, SINA will use electronic identification technologies that are already in use in different countries.

Manufacturers provide access to their private cloud solutions and devices in a well-defined way and guarantee data security and data protection using SINA. Only data and control options for which the customer has granted access will be released. The registration process alone does not release any data but signals a potential service provider the availability of certain devices and the willingness of the customer to participate.

customers define which data they grant access to, and service providers announce their requirements on data to offer services. The latter includes the type of data but also the required quality of the data, e.g. in terms of real-time requirements. The matchmaking phase succeeds when data released by a customer meet the requirements of a service provider. The SINA system shows such matches to the customer who then chooses a service provider and explicitly executes his or her rights as data owner to release the data to that explicit service for the defined functionality.

SINA will use distributed ledger technology and smart contracts to establish the contract between the customer and the service provider, so that this contract is unalterable and traceable. Thus, after the identification process, this is a further element of trust management offered by SINA.

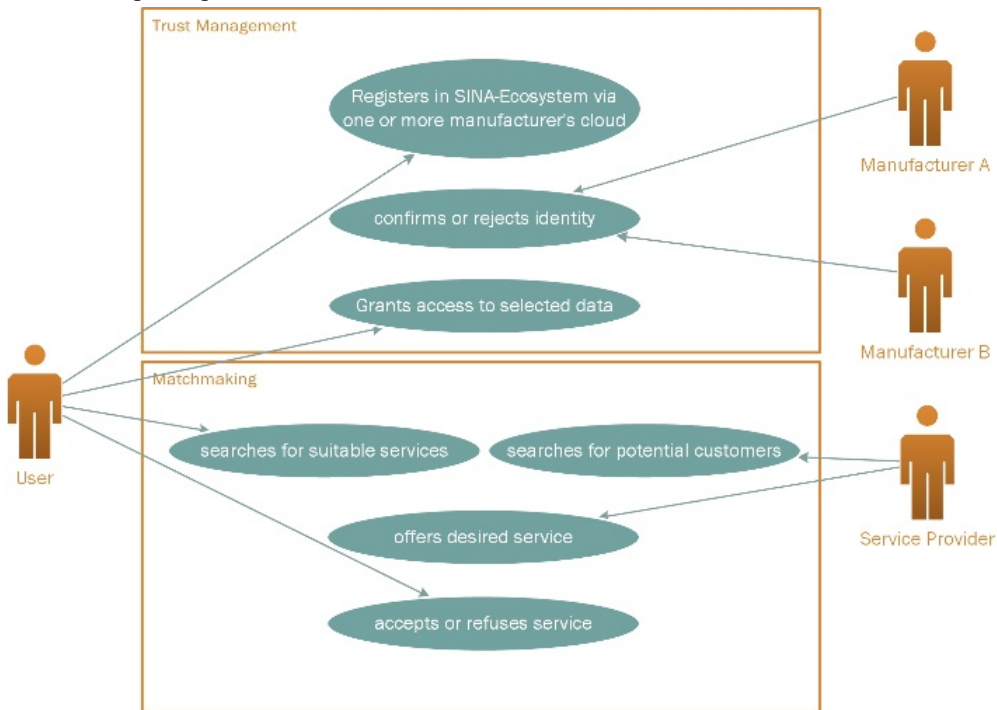

*Figure 7: Preparation and Matchmaking phase.*

### 4.4 Matchmaking phase

Due to the potentially large amount of service providers and customers, the system implements a decentralized matchmaking algorithm which finds suitable pairs of service providers and customers. The matchmaking algorithm was defined in section 4.2.3. In this phase,

## 4.5 Operational phase

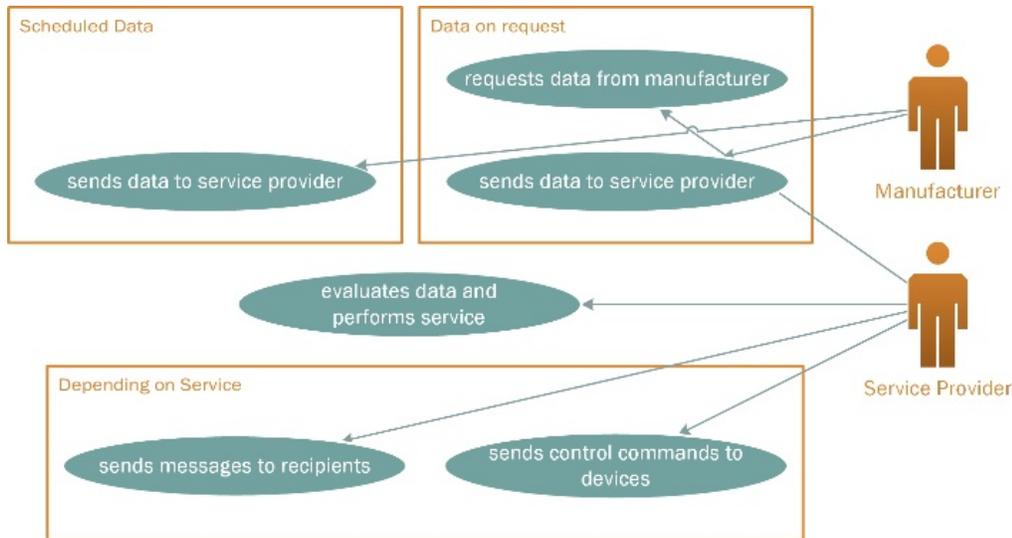

*Figure 8: Operational phase.*

After setting up the contract, the service provider gets data from the devices in the smart buildings and/or smart grid domains on request or on a scheduled base, Figure 8. It then evaluates the data to perform the service. Depending on the service, recipients get messages or devices get control commands. Messages to recipients may be all kind of messages, for example simple notifications about critical states or invoices for energy usage. Data and control commands are encrypted to ensure data integrity. If a service requires traceability and inalterability, SINA will offer these based on distributed ledger technology.

### 4.6 SINA as an IoT architecture

The proposed draft architecture is not a replacement for classical IoT architectures but is a complement to them. SINA summarizes various systems, each representing an IoT system, provided by its manufacturer, and makes the single systems transparent to service providers offering their services based on the data provided by the underlying systems. Therefore, manufacturers and service providers need not to decide to which system they want to implement in their products and services. They have the freedom to choose the most appropriate architecture and system for their requirements and can add interoperability by integrating the SINA modules.

SINA defines security and privacy related requirements and provides corresponding mechanisms. It defines and integrates functionality for registration, data management and data exchange and uses the concept of smart contracts and distributed ledger technology.

SINA architecture enables to connect things from different worlds like smart building and smart grid and makes the things interoperable. SINA makes the things and services discoverable by providing a matchmaking. The result is an IoT system with supporting functionality and a high level of security and privacy.

### 4.7 Integration of existing approaches into SINA

To integrate existing approaches, SINA relies on various levels of integration ranging from using technologies to adapting to protocols. Existing solutions can integrate the required SINA modules to become SINA-compliant. For a set of existing protocols, see the Appendix (chapter 10).

#### 4.7.1 Using technology

Particularly for blockchain and smart contracts, SINA intends to rely on already established solutions. Such solutions should already be introduced to market and have a certain prevalence. We consider this necessary in order to achieve a high level of acceptance for SINA right from the start. In addition, all components of the architecture must be evaluated to determine whether their functionality can be realized with existing technology. We will develop software or alternatively adapt the most suitable available technology only for functionalities which are not covered by existing technology.

#### 4.7.2 Adapt to protocols

In SINA we do not aim towards the development of new protocols: Several suitable protocols exist (see Appendix, chapter 10). As an example, Smart Grid Ready can be well integrated to SINA through adapters for protocols (see Figure 5 and Figure 6). The architecture will allow several adapters to be integrated simultaneously to gain the highest possible compatibility with existing systems in the area of smart buildings and smart grids.

## 4.8    Existing products

Existing products are, in the sense of SINA, manufacturer or service provider solutions. As an example, Tiko is a product connecting several devices in households through local gateways and providing a virtual power plant. To integrate Tiko into SINA means to implement the required SINA modules in the Tiko solution. Tiko then continues to offer its original service and, additionally, can add more business opportunities using the potential of the SINA ecosystem.

## 5    Application of SINA to the "Ancillary Services" sample use case

In this section, we pick up again the use case "Ancillary Services" illustrated in section 3.4. and elaborate on how SINA facilitates its implementation.

Within the use case, we have several IoT devices connected to the cloud of manufacturers and service providers. The IoT devices are in this case the EVs, charging points and the smart meters in the homes of EV owners. The DSOs, CPOs, car sharing companies and EV manufacturers have their own cloud solutions.

Each of the Information Technology and Operation Technology systems involved runs the SINA modules. In the preparation phase, each stakeholder registers with SINA, so that the identities can be trusted. Each data provider provides a semantically structured description of the type and quality of the available or needed data:

- The charging points share the charging power in real time.
- The car sharing company shares the schedule containing the bookings of their fleet.
- The car manufacturer shares the charge status and characteristics of the batteries.
- The EV owner shares forecasts or intentions about the planned EV use in the next few hours or days and whether the charger / EV support bidirectional energy flux.
- The flexibility aggregator advertises that an ancillary service can be made available by using the data above.
- The DSO shares data concerning the flexibility requests.

In the matchmaking phase, SINA recognizes that the stakeholders above can be brought together to provide a service and added value. Each EV owner and car sharing company who advertise having bidirectional charging can be matched with the flexibility aggregator and with the DSO.

Whenever a group of EV owners or a car sharing provider accept to execute a contract with the flexibility aggregator, a smart contract is generated, if possible, by picking from existing templates provided by the aggregator itself. Within the smart contract, the code which extracts the necessary data is stored together the algorithms which decide when to extract energy from a certain vehicle, together with the financial conditions or the remuneration for the involved parties. The smart contract is stored on the blockchain or distributed ledger so that it becomes immutable and traceable. If a contract is terminated, the involved party can then shut down the data access to its cloud or edge device.

In the operational phase, the smart contract is executed so that the data coming from various vehicles, the signals coming from the DSO and the schedules or use forecasts for the EVs are queried by the aggregator through SINA adapters or SINA APIs. The aggregator runs the planning algorithm contained in the smart contract to create real-time signals to be sent to the charging points or to the BMS to enact the desired charging / discharging profiles. These signals are also sent through SINA adapters or APIs. The monetary transactions which emerge during the execution are stored on the blockchain to guarantee a correct and transparent invoicing for all the involved parties.

As new EVs become available or car sharing companies switch to an EV fleet, new stakeholders participate in the matchmaking process dynamically, allowing them to profit from the services.

## 6      Conclusions and outlook

The SINA architecture and the concepts related to it are a promising way to bind the worlds of smart buildings, mobility and energy grids together to provide new business possibilities. SINA goes beyond the state of the art by providing at the same time interoperability and automatic ways to find matches between stakeholders, as well as using the blockchain and smart contracts to provide a trustworthy environment where contracts are immutable and traceable. Furthermore, the SINA components are implemented in existing edge / fog / cloud infrastructures without the need of extra hardware deployments.

As one of the next steps, we will elaborate the detailed SINA architecture including the analysis of existing standards and protocols. Those with relevance to the ecosystem will be integrated. The midterm goal is to provide a reference implementation as open source to facilitate the dissemination. In parallel we develop a governance model, which we regard as necessary to lead SINA to success by defining how the SINA ecosystem develops in future and how it is kept open for every interested partner.

The matchmaking building block will be implemented by relying on and extending existing ontologies, such as the SAREF [12] and creating an appropriate reasoning engine which can find sets of customers, service providers and manufacturers which can be combined to provide a service. Furthermore, we implement a first set of SINA adapters and the blockchain-based smart contracts engine, which will allow the first roll-out.

With the reference implementation we will start first tests in a laboratory setting to prove the feasibility of the SINA system. Together with partnering industrial companies, utilities and associations, the business models for these tests will be defined and then implemented. The implementation requires the participants to implement the SINA modules into their infrastructure, e.g., the manufacturer for household appliances must integrate the corresponding SINA adapter. Any of the partners may act as the service provider in this test phase, e.g., a utility or an academic partner.

Together with industrial partners we will test and verify the SINA system in pilot and demonstration (P+D) projects. We plan to start by implementing single business ideas. As an example, such a business case could be an energy trading platform for prosumers, comparable to the project *Quartierstrom* in Walenstadt, Switzerland [13], [14]. For a service provider offering the trading platform, SINA will foster the realization of the business model by providing access to the required devices and data through the SINA system. The service provider therefore benefits from existing infrastructure. In this case the photovoltaic systems connect to the cloud of the respective manufacturer, and the utility provides access to the smart meters. Further, the service provider can focus on the development of his service. Data security, privacy and trust management are handled by the SINA system. The P+D projects shall prove the feasibility in larger scale applications.

SINA will not only cover energy-related business models but also such for elderly care or comfort. The choice of business models will depend on the partners in the SINA ecosystem. The governance model will ensure the further development of the SINA system so that new partners with new business ideas can start their participation in the ecosystem at any time.

## 7      Ethical approval

The focus group discussions focused on the participants' expectations regarding interoperability. We did not collect personal data, nor did we carry out studies on humans. Therefore, an ethical review and approval was not required, written informed consent is not applicable.

## 8      Author contribution

The main authors are Andreas Rumsch and Christoph Imboden. They both are the drivers behind SINA and developed the architecture presented in this article. Edith Birrer, Martin Camenzind, Alberto Calatroni and Andrew Paice assisted and advised throughout the project and writing. All authors have read and agreed to the published version of the manuscript.

## 9    Acknowledgements

The authors want to express their thankfulness to those responsible in the *Innovationspark Zentralschweiz – Building Excellence*. They enabled the development of the first drafts of an architecture for SINA by providing a relevant network of companies and their representatives from which we have assembled the working group. The discussions with the members of the working group have been very inspiring and helpful to find an approach to an architecture that suits the needs of industry. The following companies are represented in the working group and support SINA. Utilities: EWL, WWZ/allthisfuture. Industry: Zehnder Group, dormakaba, Siemens Smart Infrastructure, bonainvest. Associations: Smart Grid Ready. Research: Lucerne University of Applied Sciences and Arts, CSEM.

Lucerne University of Applied Sciences and Arts allowed us to pick up and work on the idea of a smart interoperable architecture by providing the time needed to develop the idea.

## 10    Appendix: Existing Protocols or Technologies

In this section we present a collection of relevant protocols, technologies, ontologies, and associations which will be considered when implementing the concrete embodiment of the SINA architecture.

### 10.1    Relevance of existing approaches

#### 10.1.1 Associations

Often associations define standards, protocols or languages for a specific domain or use case. SINA intends to foster a variety of business models each with its use case. Therefore, SINA seeks to integrate what associations have developed. We see the following associations with complementary activities as relevant for SINA:

**EEBUS** is a non-profit association [15]. The aim is to create a uniform language standard for energy communication that enables all home devices networked with the system to communicate with each other.

**Smart Grid Ready** is a non-profit association striving to enable a smart and successful energy world that communicates [16]. The association develops a protocol-independent interoperability layer (ontology) for communication related to energy flexibility. The SmartGridReady label shall become an internationally accepted quality label. For this purpose, EEBUS and SmartGridready cooperate at different levels. It is expected that Smart Grid Ready concepts can be integrated in SINA. A Smart Grid Ready representative participates in the SINA team.

**SGReady** is a label to indicate that a Heat Pump is enabled to select 4 operation modes based on two digital input lines [17].

**VHP Ready** is a non-profit industry alliance with the aim to develop an industry standard to encourage the networking of decentralized energy systems, the certification program, as well as the appropriate testing tools in a cooperative and transparent manner together with the alliance members [18].

**Open ADR Alliance** is an industry alliance [19]. Industry stakeholders intend to build the foundation of technical activities to support the development, testing and deployment of commercial OpenADR and facilitate its acceleration and widespread adoption from it. It provides an open, highly secure, and two-way information exchange model and a global smart grid standard. It standardizes the message format used for Auto-DR and distributed energy resources management so that dynamic price and reliability signals can be exchanged in a uniform and interoperable fashion among utilities, ISOs, and energy management and control systems.

**Open Charge Alliance** (OCA) is an industry alliance of EV charging hardware and software vendors fostering global development, adoption, and compliance of the Open Charge Point Protocol and the establishment of related standards through collaboration, education, testing and certification [20].

**Smart Building Alliance for Smart Cities** (SBA) is an industry alliance for smart buildings organizing the promotion of the smart buildings sector within smart cities by associating a group of professionals [21].

**Energy Web** (EW) is a global non-profit organization that aims to use blockchain and other decentralized technologies to foster low-carbon, customer-centric electricity systems. EW focuses on building a core infrastructure and a shared technology. EW provides a toolkit to develop distributed apps using their energy web decentralized operating system [22]. The toolkit contains several building blocks that might be relevant for SINA. Particularly, it contains a Ethereum-based blockchain implementation and is able to perform smart contracts.

**G3-PLC Alliance** is a German based alliance with the aim to support, promote and implement G3-PLC in smart grid applications [23].

**Universal Smart Energy Framework** (USEF) consists of seven key players that are active across the smart energy industry, with a shared goal: one integrated smart energy system that benefits all stakeholders, from energy companies to consumers founded USEF [24]. USEF Foundation manages USEF's ongoing development. The foundation is a dedicated core team tasked with coordinating expertise, projects and partners while safeguarding the integrity and objectives of USEF.

### 10.1.2 Protocols, guidelines

SINA does not develop protocols or guidelines when they already exist. We have identified several protocols and guidelines that might be relevant for SINA. In further work, with a deeper analysis of each, we will identify those that should be implemented in SINA.

For Smart Meter Gateways the **technical guideline TR-03109** from German "Bundesamt für Sicherheit in der Informationstechnik" (BSI) defines requirements and technical specifications to obtain a secure exchange and processing of data [25, p. 03109].

**The Common Data Security Architecture** (CDSA) is a set of layered security services and cryptographic frameworks that provides an infrastructure for creating cross-platform, interoperable, security-enabled applications for client-server environments [26]. CDSA covers all the essential components of security capability, to equip applications for electronic commerce and other business applications with security services that provide facilities for cryptography, certificate management, trust policy management, and key recovery.

**The Smart Appliances REFerence** (SAREF) ontology is a shared model of consensus that facilitates the matching of existing assets (standards/protocols/data models/etc.) in the domain of smart appliances [27]. The SAREF ontology provides building blocks that allow separation and recombination of different parts of the ontology depending on specific needs.

**The Web of Things** (WoT) intends to enable interoperability across IoT platforms and application domains. WoT wants to reach this by using existing protocols and defining a set of building blocks and software patterns. The core concept of WoT is the exposure of a machine interpretable description for each thing. The descriptions cover behaviour, interaction affordances, data schema, security configuration and protocol bindings [28].

**Project Haystack** is an open source initiative that helps simplify to work with data from the Internet of Things by standardizing semantic data models and web services [29]. Using it will make it easier to unlock value from data produced by smart devices. Target applications include automation, control, energy, HVAC, lighting, and other environmental systems.

**Manufacturer Usage Description** (MUD) is an embedded software standard defined by the IETF that allows IoT device makers to advertise device specifications, including the intended communication patterns for their device when it connects to the network [30]. The network can then use this intent to author a context-specific access policy, so the device functions only within these parameters. In this manner, MUD becomes the authoritative identifier and enforcer of policy for devices on the network.

### 10.1.3 Technologies

Several technologies might be relevant for SINA. The following list of technologies describes some of the newer concepts and makes no claim to completeness in the sense that these are all the relevant technologies.

**Message Queuing Telemetry Transport** (MQTT) is a client server publish/subscribe messaging transport protocol [31]. It is lightweight, open, simple, and designed such as to be easy to implement. These characteristics make it ideal for use in many situations, including constrained environments such as communication in Machine to Machine (M2M) and Internet of Things (IoT) contexts where a small code footprint is required and/or network bandwidth is at a premium.

In the context of blockchain technology the concept of **Smart Contracts** has been created. A smart contract is a computer protocol designed to facilitate, verify or enforce digitally the negotiation or performance of a contract [32]. Smart contracts enable the execution of credible transactions without third parties. These transactions are traceable and irrevocable.

The **Blockchain technology** itself will play an important role for SINA because blockchains are designed to operate decentralized and without the need of a trusted authority [33]. Blockchains rely on permanent storage of data; also, critical, personalized data that can never be completely erased. This interferes at least partially with the general data protection regulation (GDPR).

## 11    References

[1] J. Swetina, G. Lu, P. Jacobs, F. Ennesser, and J. Song, 'Toward a standardized common M2M service layer platform: Introduction to oneM2M', *IEEE Wirel. Commun.*, vol. 21, no. 3, pp. 20–26, Jun. 2014, doi: 10.1109/MWC.2014.6845045.

[2] 'oneM2M  -  Published  Drafts'. https://www.onem2m.org/technical/published-drafts (accessed Jun. 04, 2020).

[3] 'Resources | Internet of Things World Forum'. https://www.iotwf.com/resources/71 (accessed Jun. 04, 2020).

[4] S. Cirani *et al.*, 'A Scalable and Self-Configuring Architecture for Service Discovery in the Internet of Things', *IEEE Internet Things J.*, vol. 1, no. 5, pp. 508–521, Oct. 2014, doi: 10.1109/JIOT.2014.2358296.

[5] Z. Qin, G. Denker, C. Giannelli, P. Bellavista, and N. Venkatasubramanian, 'A Software Defined Networking architecture for the Internet-of-Things', in *2014 IEEE Network Operations and Management Symposium (NOMS)*, Krakow, Poland, May 2014, pp. 1–9. doi: 10.1109/NOMS.2014.6838365.

[6] H.-C. Hsieh and C.-H. Lai, 'Internet of Things Architecture Based on Integrated PLC and 3G Communication Networks', in *2011 IEEE 17th International Conference on Parallel and Distributed Systems*, Tainan, Taiwan, Dec. 2011, pp. 853–856. doi: 10.1109/ICPADS.2011.73.

[7] J. Li, Y. Zhang, Y. Chen, K. Nagaraja, S. Li, and D. Raychaudhuri, *A Mobile Phone based WSN Infrastructure for IoT over Future Internet Architecture 1*.

[8] J. Zhou *et al.*, 'CloudThings: A common architecture for integrating the Internet of Things with Cloud Computing', in *Proceedings of the 2013 IEEE 17th International Conference on Computer Supported Cooperative Work in Design (CSCWD)*, Whistler, BC, Canada, Jun. 2013, pp. 651–657. doi: 10.1109/CSCWD.2013.6581037.

[9] D. Hanes, G. Salgueiro, P. Grossetete, R. Barton, and J. Henry, *IoT Fundamentals: Networking Technologies, Protocols, and Use Cases for the Internet of Things*. ciscopress, 2017. Accessed: Dec. 09, 2018. [Online]. Available: http://www.ciscopress.com/store/iot-fundamentals-networking-technologies-protocols-9781587144561

[10] A. Yousefpour *et al.*, 'All one needs to know about fog computing and related edge computing paradigms: A complete survey', *J. Syst. Archit.*, vol. 98, pp. 289–330, Sep. 2019, doi: 10.1016/j.sysarc.2019.02.009.

[11] CEN_CENELCE_ETSI Smart Grid Coordination Group, 'CEN-CENELEC-ETSI Smart Grid Coordination Group Smart Grid Reference Architecture'. Nov. 2012. Accessed: Jun. 09, 2021. [Online].  Available: https://ec.europa.eu/energy/sites/ener/files/documents/xpert_group1_reference_architecture.pdf

[12] 'SAREF Portal'. https://saref.etsi.org/ (accessed Jun. 09, 2021).

[13] L. Ableitner, A. Meeuw, S. Schopfer, V. Tiefenbeck, F. Wortmann, and A. Wörner, 'Quartierstrom -- Implementation of a real world prosumer centric local energy market in Walenstadt, Switzerland', *ArXiv190507242 Cs*, Jul. 2019, Accessed: Jun. 07, 2020. [Online]. Available: http://arxiv.org/abs/1905.07242

[14] A. Brenzikofer, A. Meeuw, S. Schopfer, A. Wörner, and C. Dürr, *Quartierstrom: A Decentralized Local P2P Energy Market Pilot On A Self-Governed Blockchain*. AIM, 2019. Accessed: Jun. 07, 2020. [Online]. Available: https://www.cired-repository.org/handle/20.500.12455/181

[15] EEBus Initiative e.V., 'EEBus Whitepaper 2.0.pdf', *http://www.eebus.org*, 2013. https://www.eebus.org/fileadmin/Mediapool/Download/downloads_en/2013_08_EEBus_e_V_Whitepaper_2.0__e_.pdf (accessed Mar. 14, 2014).

[16] 'Smart Grid Ready | Schweiz | Home', *SMART GRID READY*. https://www.smartgridready.ch (accessed Jun. 07, 2020).

[17] 'SG  Ready-Label'. https://www.waermepumpe.de/normen-technik/sg-ready/ (accessed Jun. 07, 2020).

[18] 'Der Kommunikationsstandard für Smart Grids. - VHPready  Services  GmbH'. https://www.vhpready.de/de/home/ (accessed Jun. 07, 2020).

[19] 'OpenADR Alliance'. https://www.openadr.org/ (accessed Jun. 07, 2020).

[20] 'Open  Charge  Alliance'. https://www.openchargealliance.org/ (accessed Jun. 07, 2020).

[21] 'Smart Builgins Alliance for Smart Cities', *Smart Buildings  Alliance*. https://www.smartbuildingsalliance.org/en/home (accessed Jun. 07, 2020).

[22] N. Gavhane *et al.*, 'EnergyWeb-EWDOS-VisionPurpose-vFinal-20200309.pdf'. Dec. 2019. Accessed: Apr. 21, 2020. [Online]. Available: https://www.energyweb.org/wp-

content/uploads/2019/12/EnergyWeb-EWDOS-VisionPurpose-vFinal-20200309.pdf

[23] 'G3-PLC Alliance Home'. http://www.g3-plc.com/home/ (accessed Jun. 07, 2020).

[24] 'Usef Energy – Universal Smart Energy Framework'. https://www.usef.energy/ (accessed Jun. 07, 2020).

[25] Bundesamt für Sicherheit in der Informationstechnik (BSI), 'Technische Richtlinie BSI TR-03109-1 Anforderungen an die Interoperabilität der Kommunikationseinheit eines intelligenten Messsystems', Bonn, Version 1.0.

[26] 'Common Security: CDSA and CSSM, Version 2 (with corrigenda)'. https://publications.opengroup.org/c914 (accessed Jun. 07, 2020).

[27] S. Dahmen-Lhuissier, 'ETSI - Smart Appliances - Smart Domestic & Industrial Appliances', *ETSI*. https://www.etsi.org/technologies/smart-appliances (accessed Jun. 07, 2020).

[28] 'W3C Web of Things at W3C'. https://www.w3.org/WoT/ (accessed Jun. 07, 2020).

[29] 'Project Haystack - Home', *Project Haystack*. http://project-haystack.org/ (accessed Dec. 15, 2016).

[30] E. Lear, 'Manufacturer Usage Description Specification', May 25, 2018. https://tools.ietf.org/id/draft-ietf-opsawg-mud-22.html (accessed Jun. 07, 2020).

[31] 'MQTT'. https://mqtt.org/ (accessed Jun. 07, 2020).

[32] P. D. A. Mitschele, 'Definition: Smart Contract', *https://wirtschaftslexikon.gabler.de/definition/smart-contract-54213*. https://wirtschaftslexikon.gabler.de/definition/smart-contract-54213 (accessed Jun. 07, 2020).

[33] P. D. A. Mitschele, 'Definition: Blockchain', *https://wirtschaftslexikon.gabler.de/definition/blockchain-54161*. https://wirtschaftslexikon.gabler.de/definition/blockchain-54161 (accessed Jun. 07, 2020).